\documentclass{moriond}

\usepackage{amsmath}
\usepackage{amssymb}

\def\pa{\partial}

\def\ov{\overline}

\def\be{\begin{equation}}
\def\ee{\end{equation}}
\def\bea{\begin{eqnarray}}
\def\eea{\end{eqnarray}}

\newcommand{\Photo}{}

\begin{document}
\vspace*{4cm}
\title{EXPERIMENTAL SIGNATURES \\OF A LIGHT SINGLET-LIKE SCALAR IN
   NMSSM}

\author{ MARCIN BADZIAK\footnote[1]{Speaker}, MAREK OLECHOWSKI, STEFAN POKORSKI}

\address{Institute of Theoretical Physics, Faculty of Physics, University of Warsaw\\
        ul.\ Ho\.za 69, PL--00--681 Warsaw, Poland}

\maketitle

\abstracts{
NMSSM with a light singlet-like scalar and strongly suppressed
couplings to $b$ and $\tau$ is investigated.
It is shown that in such a scenario the singlet-like scalar to diphoton signal
can be larger than for the SM Higgs for a wide range of masses between 60 and
110 GeV, in agreement with all the LEP and LHC data. Enhancement of the
singlet-like scalar to diphoton signal is correlated with positive correction to
the SM-like Higgs mass from mixing between SM-like Higgs and the singlet. It is
also shown that the couplings to $b$ and $\tau$ and, in consequence, branching
ratios of the SM-like Higgs are anti-correlated with those of the singlet-like
scalar. 
If the singlet-like scalar to diphoton signal is enhanced, the signal strengths
of the 125 GeV Higgs in the diphoton and $WW^*/ZZ^*$ channels are predicted to
be smaller than for the SM Higgs.}

\section{Introduction}

The measurement of the Higgs mass of about 125 GeV \cite{Atlas_discovery,CMS_discovery} stimulated increased interest in extensions of 
Minimal Supersymmetric Standard Model (MSSM) in which additional positive corrections to the Higgs mass are present 
allowing for lighter stops relaxing the little hierarchy problem of the MSSM.
One of the most attractive and certainly the simplest among those extensions is Next-to-Minimal Supersymmetric Standard Model (NMSSM). 
In these proceedings we focus on a case in which additional positive correction to the Higgs mass originates from mixing effects in the CP-even scalar
sector of NMSSM.
We identify particularly attractive part of parameter space in which the singlet-light scalar is lighter than the discovered Higgs and has
suppresed couplings to $b$ and $\tau$. In this part of parameter space more than 5 GeV correction to the MSSM Higgs mass is obtained for wide range
of singlet-like scalar masses. Such scenario has phenomenologically distinctive feature which is diphoton signal of the singlet-like scalar stronger
than the corresponding signal of the SM Higgs with the same mass. 
The results presented in these proceedings were originally given in Refs.~\cite{nmssmmixing,nmssmmixing2}.

\section{CP-even scalar sector in NMSSM}
\label{sec2}

We begin with a brief summary of the CP-even scalar
sector of NMSSM \cite{reviewEllwanger}. There are three physical neutral CP-even
Higgs fields, $H_u$, $H_d$, $S$ which are the real parts of the excitations
around the real vevs, $v_u\equiv v \sin\beta$, $v_d\equiv v \cos\beta$, $v_s$
with $v^2=v_u^2 + v_d^2\approx (174 \ {\rm GeV})^2$,  of the neutral components of
doublets $H_u$, $H_d$ and the singlet $S$ (we use the same notation for the
doublets and the singlet as for the real parts of their neutral components). It
is more convenient for us to work in the  basis $(\hat{h}, \hat{H}, \hat{s})$,
where $\hat{h}=H_d\cos\beta + H_u\sin\beta$, $\hat{H}=H_d\sin\beta -
H_u\cos\beta$ and $\hat{s}=S$. The $\hat{h}$ field has exactly the same
couplings to the gauge bosons and fermions as the SM Higgs field. The field
$\hat{H}$ does not couple to the gauge bosons and its couplings to the up and
down fermions are the SM Higgs ones rescaled by $\tan\beta$ and $-\cot\beta$,
respectively. The mass eigenstates are denoted as $s$, $h$, $H$, with the
understanding that $h$ is the SM-like Higgs.

The NMSSM specific part of the superpotential is in general given
by\footnote{Explicit MSSM-like $\mu$-term can also be present in the
superpotential but it can always be set to zero by a constant shift of the real
component of S.}
\begin{equation}
\label{W_NMSSM}
 W_{\rm NMSSM}= \lambda SH_uH_d + f(S) \,.
\end{equation}
The first term is the source of the effective higgsino mass parameter,
$\mu\equiv\lambda v_s$,
while the second term parametrizes various versions of NMSSM. In the simplest
version, known as the scale-invariant NMSSM, $f(S)\equiv\kappa S^3/3$.

We assume also quite general pattern of soft SUSY breaking terms (we follow the
conventions used in \cite{reviewEllwanger}): 
\begin{equation}
 -\mathcal{L}_{\rm soft}\supset m_{H_u}^2 |H_u|^2 + m_{H_d}^2 |H_d|^2 + m_S^2
|S|^2 + ( A_{\lambda} \lambda H_u H_d S +\frac{1}{3}\kappa A_{\kappa} S^3 +
m_3^2 H_u H_d + \frac{1}{2}m_S^{\prime2} S^2 + \xi_S S + {\rm h.c.} )\,. 
\end{equation}
In the scale-invariant NMSSM, $m_3^2=m_S^{\prime2}=\xi_S=0$.

Let us parametrize the mass matrix of the hatted fields as follows:
\begin{equation}
 \hat{M}^2=
\left(
\begin{array}{ccc}
  \hat{M}^2_{hh} & \hat{M}^2_{hH} & \hat{M}^2_{hs} \\[4pt]
   \hat{M}^2_{hH} & \hat{M}^2_{HH} & \hat{M}^2_{Hs} \\[4pt]
   \hat{M}^2_{hs} & \hat{M}^2_{Hs} & \hat{M}^2_{ss} \\
\end{array}
\right) \,.
\end{equation}
The off-diagonal terms of the above matrix are given by 
\begin{align}
\label{MhH}
& \hat{M}^2_{hH} = \frac{1}{2}(M^2_Z-\lambda^2 v^2)\sin4\beta \,, \\
\label{Mhs}
& \hat{M}^2_{hs} =  \lambda v (2\mu-\Lambda \sin2\beta) \,, \\
\label{MHs}
& \hat{M}^2_{Hs} = \lambda v \Lambda \cos2\beta \,,
\end{align}
where $\Lambda\equiv A_{\lambda}+\langle\pa^2_S f\rangle$.

$\hat{M}^2_{hh}$ after including radiative corrections, which we parametrize by
$(\delta m_h^2)^{\rm rad}$, is given by\footnote{
The formulea for the diagonal entries of $\hat{M}^2$ are not relevant for our
present discussion and can be found in Ref.~\cite{nmssmmixing}. }
\begin{equation}
 \label{Mhh}
 \hat{M}^2_{hh} = M_Z^2\cos^2\left(2\beta\right)+(\delta m_h^2)^{\rm rad} +
\lambda^2 v^2\sin^2\left(2\beta\right) \,.
\end{equation}
The first two terms in the above equation are the ''MSSM'' terms,
with
\begin{equation}
 (\delta m_h^2)^{\rm rad} \approx \frac{3g^2m_t^4}{8 \pi^2 m_W^2}
\left[\ln\left(\frac{M_{\rm SUSY}^2}{m_t^2}\right)+\frac{X_t^2}{M_{\rm SUSY}^2}
\left(1-\frac{X_t^2}{12M_{\rm SUSY}^2}\right)\right] \,,
\end{equation}
where $M_{\rm SUSY}\equiv\sqrt{m_{\tilde{t}_1}m_{\tilde{t}_2}}$
($m_{\tilde{t}_i}$ are the stop masses)
and $X_t\equiv
A_t-\mu/\tan\beta$ with $A_t$ being SUSY breaking top trilinear coupling at
$M_{\rm SUSY}$.
The third term in eq.~(\ref{Mhh}) is the new tree-level contribution coming from
the $\lambda SH_uH_d$ coupling.

\subsection{Higgs boson couplings}

Denoting the mass-eigenstates $s$, $h$, $H$ by $x=\ov{g}_x\hat{h} +
\beta^{(H)}_x\hat{H} + \beta^{(s)}_x\hat{s}$ the couplings (normalized to the
corresponding SM
values) are given by
\begin{align}
\label{Cb}
 &C_{b_x}=\ov{g}_x+\beta^{(H)}_x\tan\beta \,, \\
 &C_{t_x}=\ov{g}_x-\beta^{(H)}_x\cot\beta \,, \\
 &C_{V_x}=\ov{g}_x \,,
\end{align}
where $x$ is $s$, $h$ or $H$. Note that the couplings to the vector bosons
depend only on the $\hat{h}$ components. On the other hand, the couplings to
fermions can be modified by mixing with $\hat{H}$.    

In the region of moderate and  large $\tan\beta$  even small component of
$\hat{H}$ in the singlet-dominated Higgs may give a large contribution to the
couplings to $b$ quark due to $\tan\beta$ enhancement. On the other hand, the
couplings to the up-type quarks are almost the same as those to the gauge
bosons, $C_{t_x} \approx C_{V_x}$. Particularly interesting is the case when
$\ov{g}_s \beta^{(H)}_s<0$ because then $C_{b_s}\ll
C_{t_s}, C_{V_s}$ is possible.

It can be shown that $C_{b_s} < C_{V_s}$ if $\hat{M}^2_{Hs} \hat{M}^2_{hs}
<0$ which leads to the following condition for the NMSSM parameters
\cite{nmssmmixing}:
\begin{equation}
 \Lambda(\mu\tan\beta-\Lambda)\gtrsim0 \,,
\end{equation}
which is satisfied only if $\mu\Lambda>0$.

It is important to stress that strong suppression of the $sb\bar{b}$ coupling
does not introduce any new fine-tuning of parameters as long as $\tan\beta$ is
large and $\lambda$ is smaller than ${\mathcal O}(0.1)$. The
scenario with large $\lambda\gtrsim0.6$ and small $\tan\beta$ that lead to
substantial tree-level contribution to $m_h$ requires at least few-percent
fine-tuning to avoid negative eigenvalues of the Higgs mass matrix
\cite{nmssmmixing}. 

In the regime $C_{b_s}\ll C_{t_s}, C_{V_s}$, the (otherwise dominating) $s$
branching ratios to $b\bar{b}$ and $\tau\bar{\tau}$ are strongly suppressed and
$s$ decays mainly to $gg$ and $c\bar{c}$. This has many important implications
which we discuss in following sections.

\section{Correction to the Higgs mass from mixing}

In NMSSM, in addition to the correction coming from
the $\lambda$ coupling, there exist also correction
from mixing between scalars, $\Delta_{\rm mix}$, which we parametrize as
follows:
\begin{equation}
 m_h=\hat{M}_{hh} + \Delta_{\rm mix} \,.
\end{equation}
Mixing with
$\hat{s}$ gives positive (negative) contribution to $\Delta_{\rm mix}$ if the
singlet-dominated scalar is lighter (heavier) then the SM-like Higgs, while
the contribution to $\Delta_{\rm mix}$ from mixing with $\hat{H}$ is negative
and gets smaller in magnitude for larger values of $m_H$. 
Therefore, in order to calculate the maximal allowed value of $\Delta_{\rm mix}$
it is enough to consider a case in which $H$ is decoupled.
In such a case $\Delta_{\rm mix}$ is given by
\begin{equation}
\label{deltamix}
 \Delta_{\rm mix} = m_h - \sqrt{m_h^2 - \ov{g}^2_s \left(m_h^2-m_s^2\right)}
\approx \frac{\ov{g}^2_s}{2} \left(m_h - \frac{m_s^2}{m_h} \right)
+\mathcal{O}(\ov{g}^4_s) \,,
\end{equation}
where in the last, approximate equality we used the expansion in
$\ov{g}^2_s\ll1$. It is clear from the above formula that a substantial
correction to the Higgs mass from the mixing is possible only for not too small
couplings of the singlet-like state to the $Z$ boson and that $m_s\ll m_h$ is
preferred. However, LEP has provided rather strong constraints on the states
with masses below $\mathcal{O}(110)$ GeV that couple to the $Z$ boson because
such states could be copiously produced in the process $e^+e^-\to sZ$.

\begin{figure}[t!]
  \begin{center}
    \includegraphics[width=0.4\textwidth]{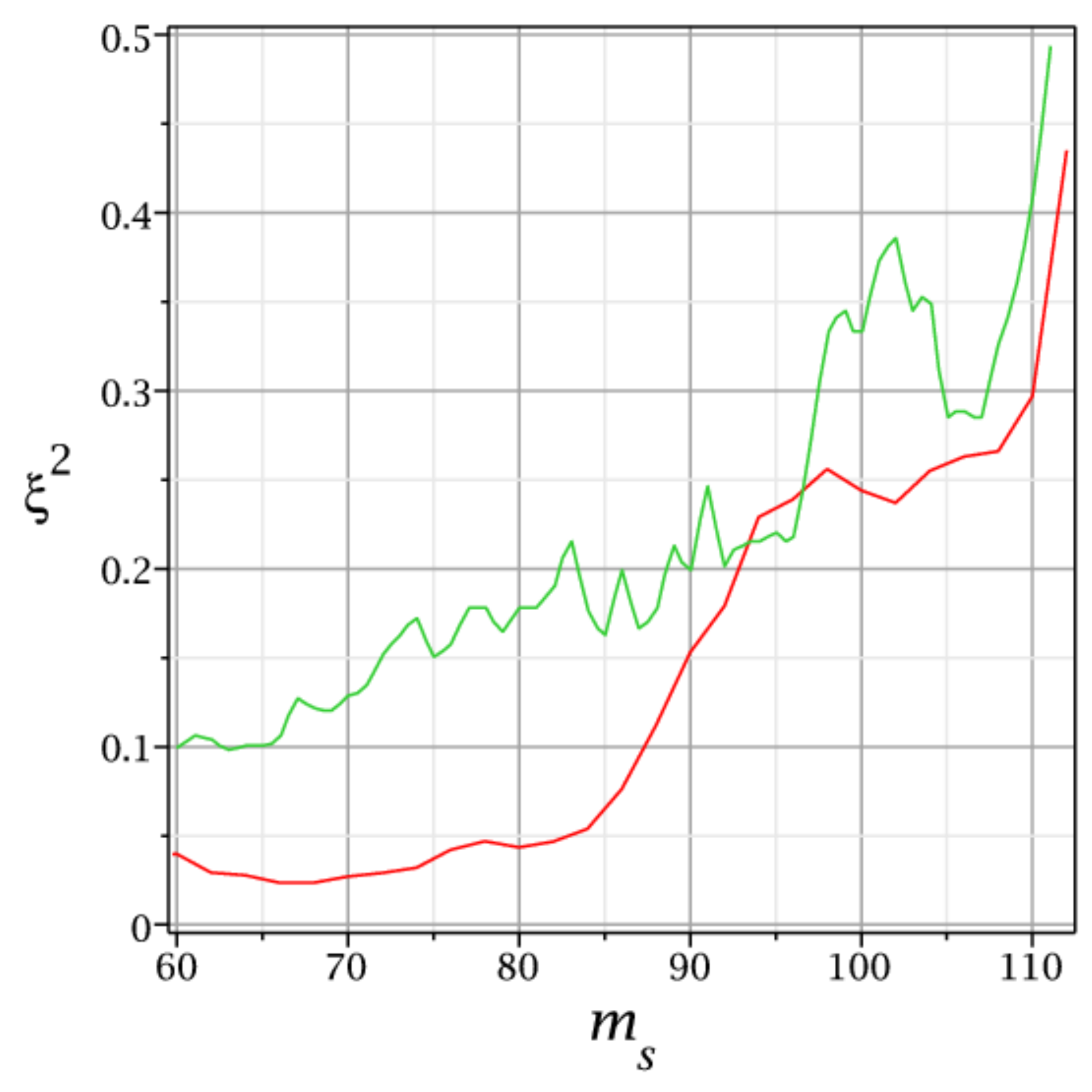}
    \includegraphics[width=0.4\textwidth]{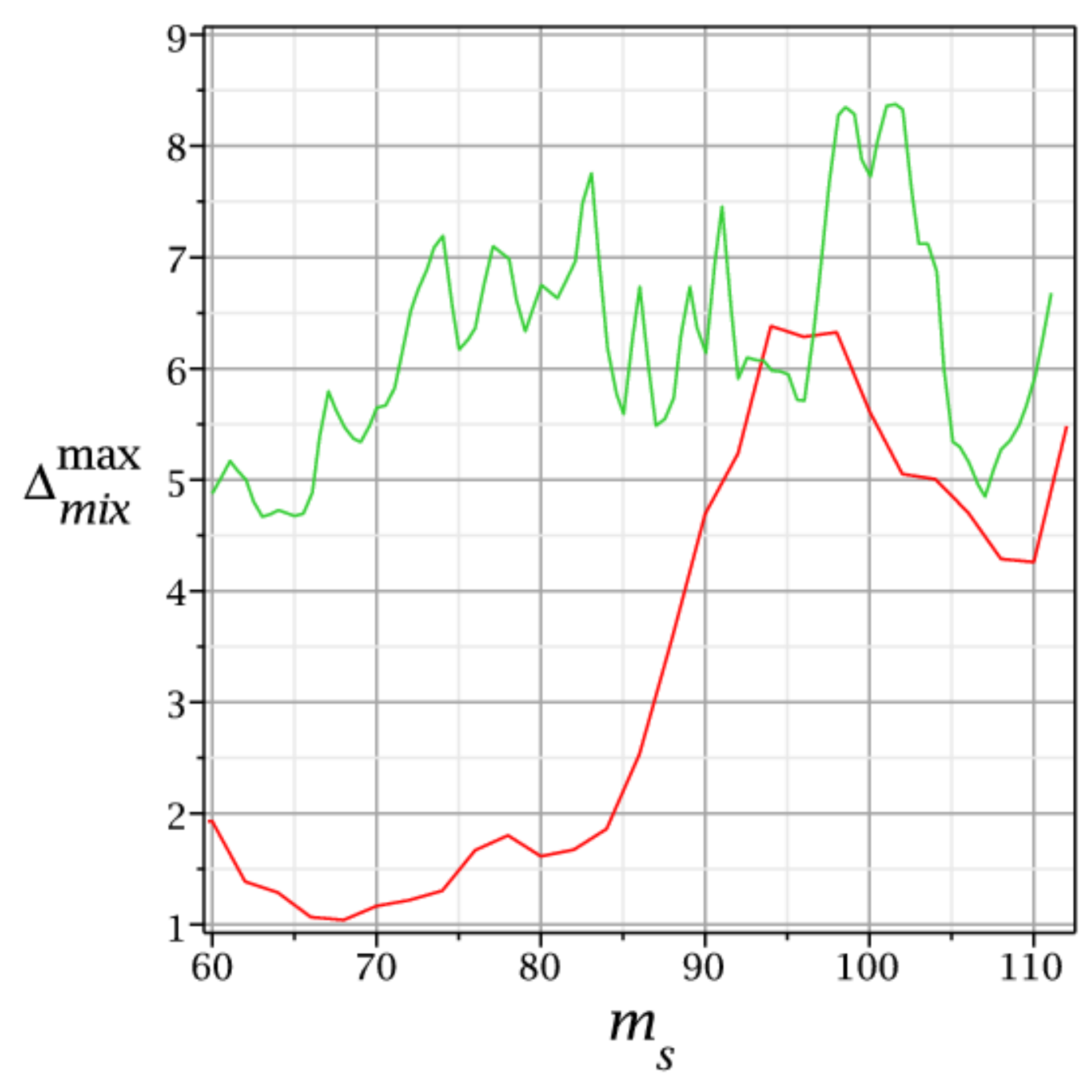}
    \caption{Left: The 95 \% CL upper bounds on $\xi_{b\ov{b}}^2$ (the red line) or $\xi_{jj}^2$ (the green line). The red line was obtained
using the observed limits presented in the column (a) of Table 14 in Ref.~\protect\cite{LEP_bb}, while the green line corresponds to Figure 2 of
\protect\cite{LEP_jj}. Right: The LEP limits translated to the upper limits on
$\Delta_{\rm mix}$ using eq.~\ref{deltamix} assuming $\xi_{b\ov{b}}^2=\ov{g}_s^2$ and $\xi_{jj}^2=\ov{g}_s^2$ for the red and green line,
respectively.}
\label{fig:deltamix_ms_LEP}
  \end{center}
\end{figure}

For those LEP searches that rely on $b$ and $\tau$ tagging \cite{LEP_bb},
constraints on $\ov{g}^2_s$ depend on the $s$
branching ratios and the LEP experiments provide constraints on the quantity
$\xi^2$ defined as:
\begin{equation}
\label{xidef}
 \xi_{b\bar{b}}^2\equiv \ov{g}^2_s \times \frac{{\rm BR}(s\to b\bar{b})}{{\rm
BR^{\rm SM}}(h\to b\bar{b})} \,.
\end{equation}
The LEP constraints on $\xi_{b\bar{b}}^2$ are reproduced by the red line in the
left panel of Figure \ref{fig:deltamix_ms_LEP}. Assuming $C_{b_s}\approx
C_{t_s}, C_{V_s}\approx \ov{g}_s$, i.e. neglecting the mixing with $\hat{H}$, 
the branching ratios of $s$ are the same as for the SM Higgs. Therefore, the
limits on $\xi_{b\bar{b}}^2$  depicted by the red line  in Figure
\ref{fig:deltamix_ms_LEP} are, in fact, also the limits on $\ov{g}_s^2$. 
Using eq.~(\ref{deltamix}) we can translate the constraints on $\ov{g}_s^2$ into
limits for the maximal allowed correction from the mixing, $\Delta_{\rm
mix}^{\rm max}$, as a function of $m_s$. These are presented in the right panel
of Figure \ref{fig:deltamix_ms_LEP}. Notice that in this case $\Delta_{\rm mix}$
can reach about 6 GeV in a few-GeV interval for
$m_s$ around 95 GeV, where the LEP experiments observed the $2\sigma$ excess in
the $b\bar{b}$ channel.
However, for $m_s\lesssim90$ GeV the allowed value of $\Delta_{\rm mix}^{\rm
max}$ drops down very rapidly to very small values.

The situation changes if the $sb\bar{b}$ coupling is strongly suppressed because in such a case
the constraints from the $b$-tagged LEP searches become meaningless. 
The main constraint on such a scenario is provided by the flavour independent
search for a Higgs decaying into
two jets at LEP \cite{LEP_jj}. 
Those searches give constraints on a quantity $\xi_{jj}^2$
defined as:
\begin{equation}
 \xi_{jj}^2\equiv \ov{g}^2_s \times {\rm BR}(s \to jj) \,,
\end{equation}
which are reproduced by the green line in the left panel of
Figure \ref{fig:deltamix_ms_LEP}. Noting that for suppressed $sb\bar {b}$ and 
$s\tau\bar{\tau}$ couplings, ${\rm BR}(s \to jj)\approx1$ so $\xi_{jj}^2\approx
\ov{g}^2_s$, it becomes clear that these constraints are substantially weaker
and allow for values
of $\ov{g}_s^2$ above 0.3 for $m_s$ around 100 GeV and the limit rather slowly
improves as $m_s$ goes down, as seen from the left panel of Figure
\ref{fig:deltamix_ms_LEP}. In consequence, the constraints on $\Delta_{\rm
mix}^{\rm max}$ are also weaker.
As can be seen from the right panel of Figure \ref{fig:deltamix_ms_LEP}, when
$s\to b\bar{b}$ decays are suppressed $\Delta_{\rm mix}$ above 5 GeV is viable
for a large range of $m_s$ with a maximum of about 8 GeV for $m_s$ around 100
GeV.

\subsection{Anti-correlation between the $s$ and $h$ couplings}

\begin{figure}[t!]
  \begin{center}
    \includegraphics[width=0.4\textwidth]{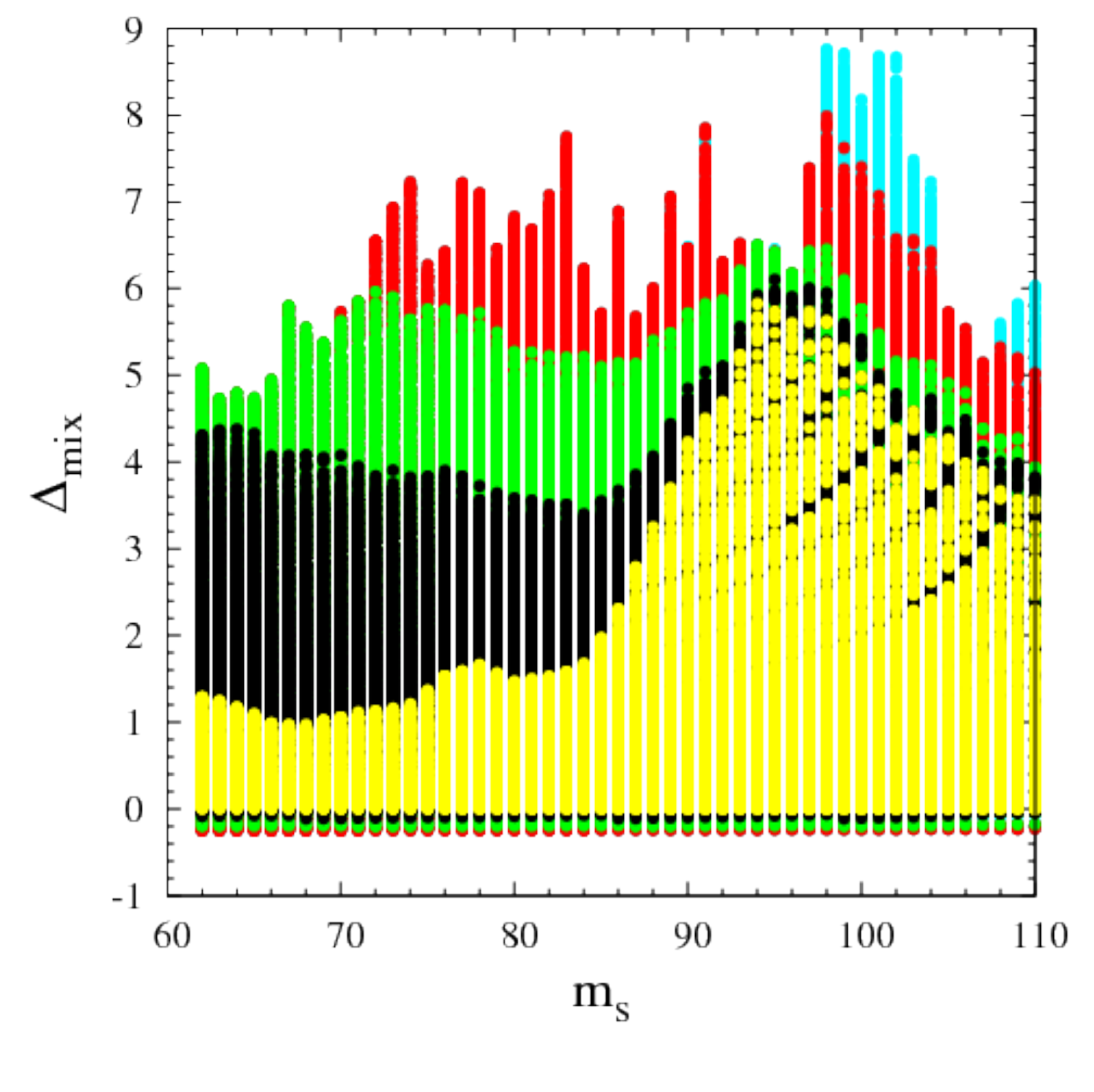}  
\includegraphics[width=0.4\textwidth]{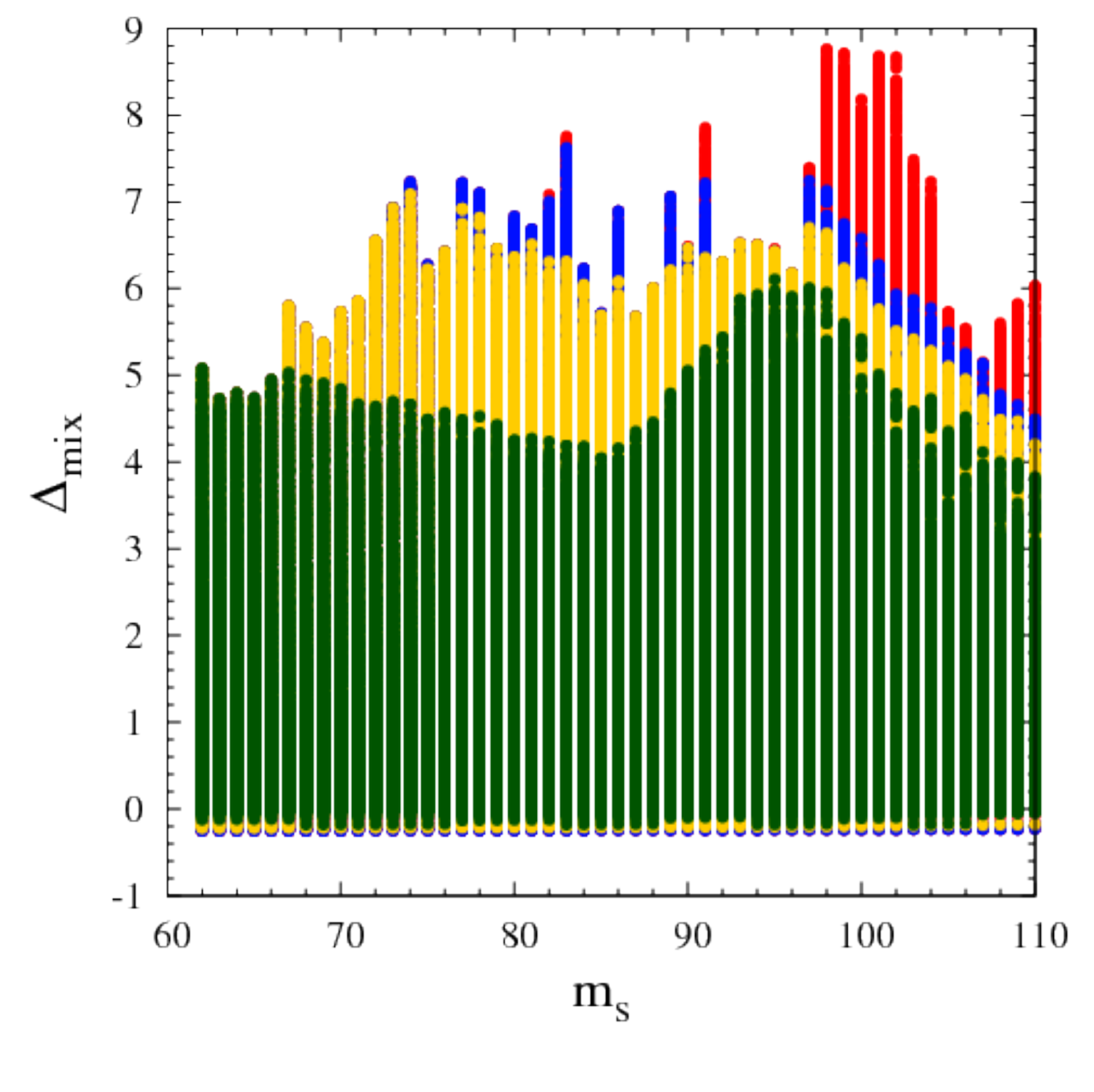}
    \caption{Results of the numerical scan presented in the $\Delta_{\rm
mix}$-$m_s$ plane after taking into account the LEP constraints. Left: The blue points are characterised by $R_{VV}^{(h)}<0.5$ while
for the red, green, black  and yellow points $R_{VV}^{(h)}$ is larger than 0.5,
0.7, 0.8 and 1, respectively.  
              The points with larger values of $R_{VV}^{(h)}$ are overlaid on
the points with smaller $R_{VV}^{(h)}$. 
Right: Color denotes $p$-value based on the 125 GeV Higgs data. 
Green, yellow, blue, red points correspond to $p$-value above 0.32, between 0.32 and 0.05, between 0.05 and 0.01, below 0.01, respectively.
    }
    \label{fig:deltamix_ms_scan}
  \end{center}
\end{figure}

It is very interesting to note that the $sb\bar{b}$ coupling is anti-correlated
with the $hb\bar{b}$ coupling. The proof of this fact is given in
Ref.~\cite{nmssmmixing}.
This implies that for suppressed $sb\bar{b}$ coupling,
$R_{VV}^{(h)}<1-\ov{g}_s^2$ (where $V=W$ or $Z$, and $R_{VV}^{(h)}$ is the cross-section times branching ratio normalized to the SM value). In order
to study
quantitatively this feature of the scenario we performed a numerical scan over
the NMSSM parameter space for various values of $m_s$ and $m_H$ while keeping
fixed $m_h=125$ GeV. 
Details of the scan can be found in Ref.~\cite{nmssmmixing} and the results are presented in Figures \ref{fig:deltamix_ms_scan} and \ref{fig:Rgam_s}.
In Figure
\ref{fig:deltamix_ms_scan} scatter plots of $\Delta_{\rm mix}$ versus $m_s$ are
presented. The LEP constraints discussed before have been taken into account. It
can be seen from the left panel that the biggest $\Delta_{\rm mix}$ corresponds to small values of $R_{VV}^{(h)}$ (note that for large $\tan\beta$
$R_{\gamma\gamma}^{(h)}\approx R_{VV}^{(h)}$), in tension with the LHC Higgs data.
If one demands consistency with the LHC Higgs data at 95\% CL, $\Delta_{\rm mix}^{\rm max}$ 
is between about 5 and 7 GeV for $m_s$ between $m_h/2$ and 105 GeV, as can be seen from the right panel of Figure \ref{fig:deltamix_ms_scan}.

\section{Enhanced $s\to\gamma\gamma$}

In the scenario with strongly supressed $sb\bar{b}$ and $s\tau\bar{\tau}$ couplings the total decay width of $s$ is strongly reduced so 
all the $s$ branching ratios, except those for the $s$ decays to the down-type fermions, are strongly enhanced. 
Particularly interesting, from the viewpoint of prospects for discovery of $s$, is the $\gamma\gamma$ final state. 
Even though the production cross-section is suppressed by $\ov{g}_s^2$, ${\rm BR}(h\to \gamma\gamma)$ can be enhanced 
by more than a factor of ten resulting in the signal strength $R_{\gamma\gamma}^{(s)}>1$.

\begin{figure}[t!]
  \begin{center}
   
\includegraphics[width=0.4\textwidth]{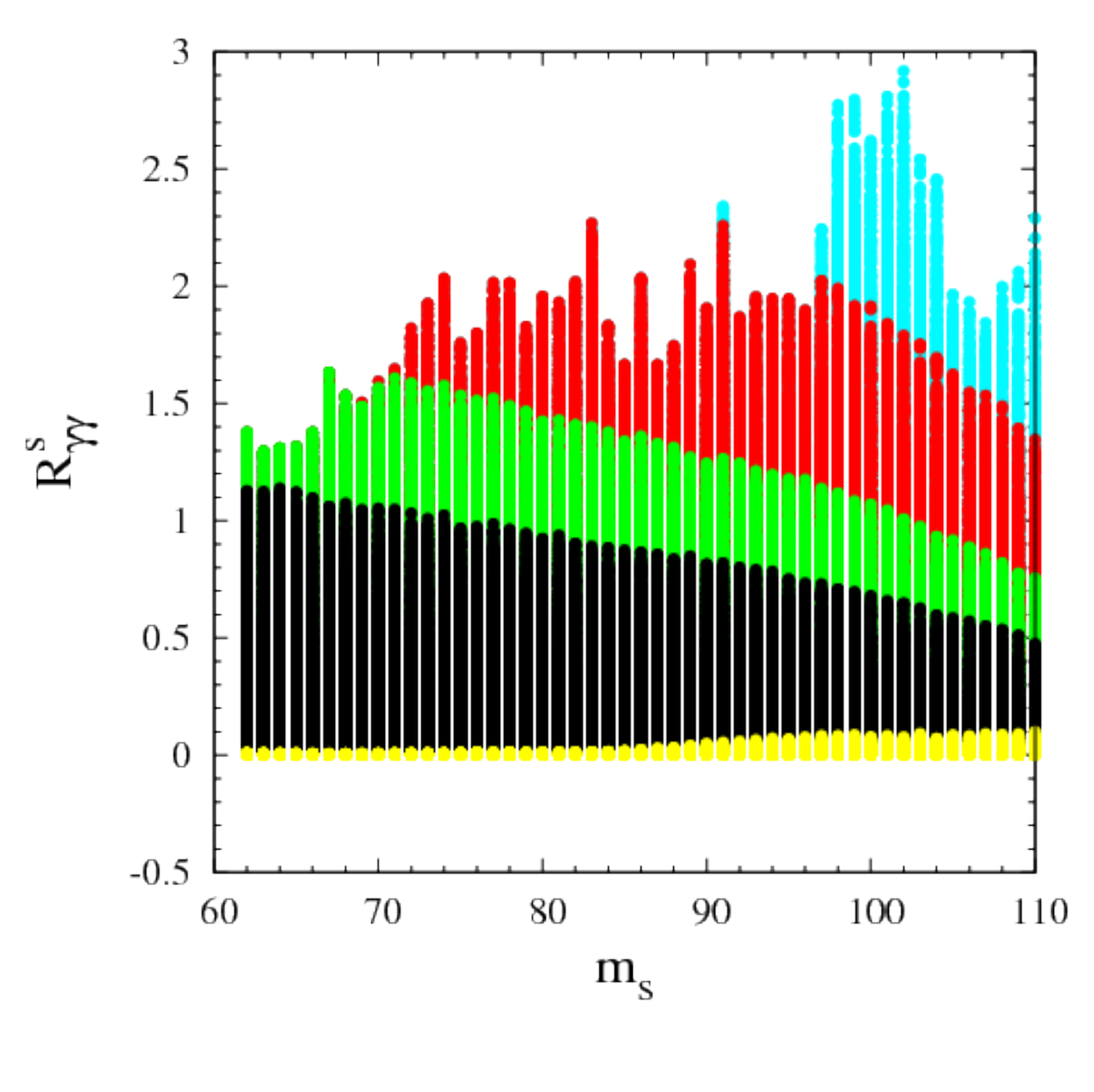}
\includegraphics[width=0.4\textwidth]{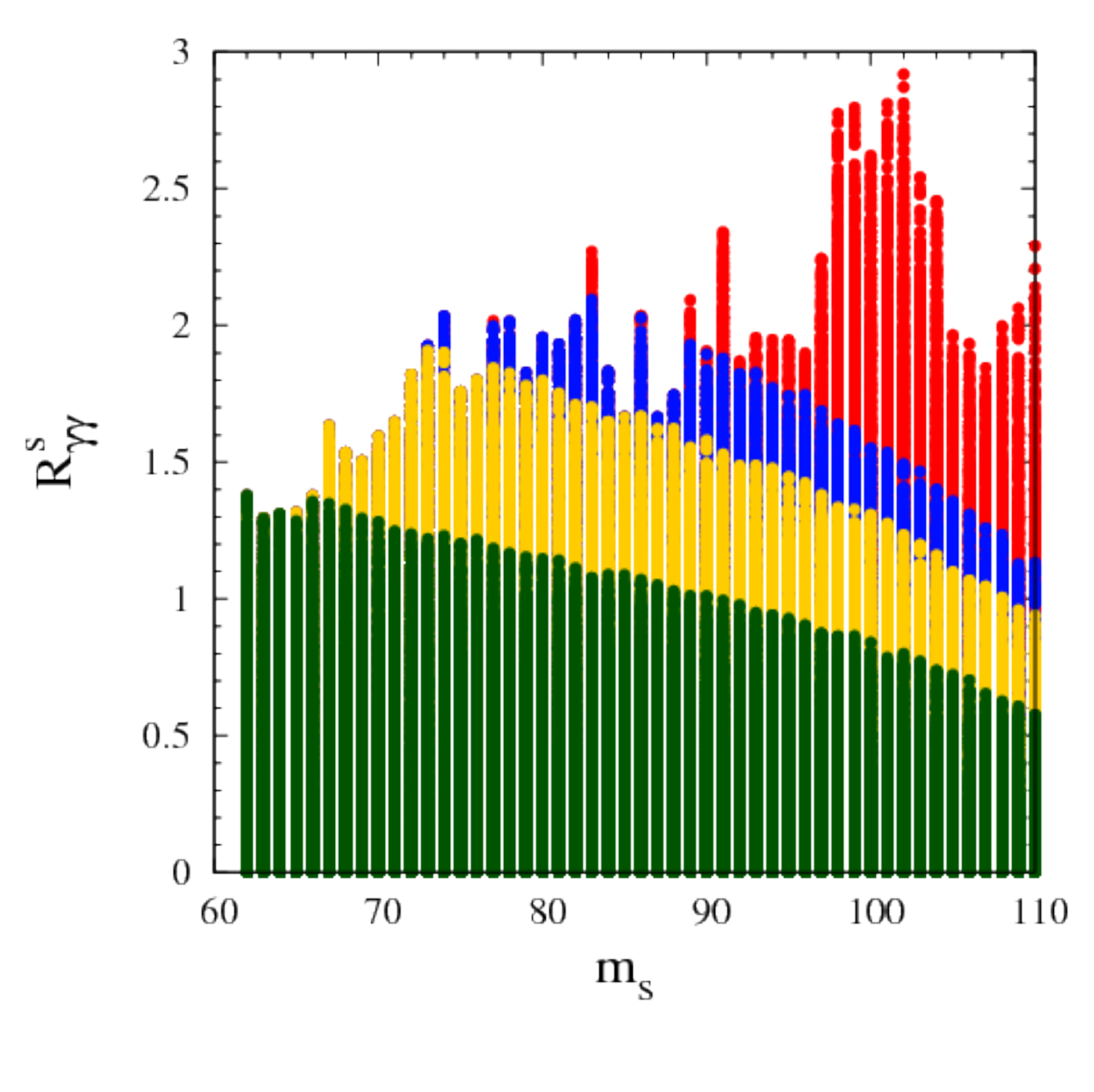}
    \caption{The same as in Figure \ref{fig:deltamix_ms_scan} but in the $R_{\gamma\gamma}^{(s)}$-$m_s$ plane. 
    }
    \label{fig:Rgam_s}
  \end{center}
\end{figure}

It can be seen from Figure \ref{fig:Rgam_s} that after taking into account the LEP constraints $R_{\gamma\gamma}^{(s)}$ can be up to about three. 
However, it is also seen from the Figure that very large $R_{\gamma\gamma}^{(s)}$ typically correspond to small $R_{VV}^{(s)}$ and is incompatible
with the LHC data for the 125 GeV Higgs.
Nevertheless, even after taking into account these constraints $R_{\gamma\gamma}^{(s)}>1$ can be obtained for a wide range of $m_s$ between about 60
and 110 GeV,
and $R_{\gamma\gamma}^{(s)}$ can reach two for $m_s$ around 70 GeV.

\section{Conclusions}

We investigated NMSSM with a light singlet-like scalar with suppressed couplings
to $b$ and $\tau$. 
In this scenario the discovered Higgs boson is next to the lightest one and its
mass can get positive correction of about 5-7 GeV from mixing with the singlet
for a wide range of $m_s$ between 60 and 110 GeV. Thus, stop masses can be
substantially smaller than in the MSSM.

Characteristic signature of this scenario is enhanced $s \to
\gamma\gamma$ signal which can be up to three times larger than for the SM Higgs
in agreement with all the LEP data. 
Even though direct Higgs searches in the diphoton
channel  have not been performed for masses below 110 GeV at the LHC so far, the
LHC results already constrain this scenario because the 125 GeV Higgs production
cross-section and branching ratios to gauge bosons are always smaller than in
the SM. This is because mixing of $\hat{h}$ with the singlet reduces the Higgs
production cross-section, while suppression of the $sb\bar{b}$ coupling always
imply enhancement of the $hb\bar{b}$ coupling.
We found that the NMSSM points predicting the largest enhancement of the $s \to
\gamma\gamma$ signal are ruled out by the 125 GeV Higgs measurements but the
signal strength two times larger than for the SM Higgs is still possible.
Therefore, we encourage the LHC collaborations to extend direct Higgs searches
in the diphoton channel to masses between 60 and 110 GeV. 

Since the enhancement of the $s \to \gamma\gamma$ signal originates from strong suppression of the total decay width, similar enhancement of signal is
present also in other $s$ decay channels  such as $WW^*$, $ZZ^*$ and $Z\gamma$. For low scalar masses these channels are generally even more
challenging from experimental point of view but for masses not much below 110 GeV the signal of $s$ decays may be visible. Thus, the Higgs searches
should be extended to lower masses also in these channels.  

Suppression of the $sb\bar{b}$ coupling is naturally present in NMSSM with
moderate and large $\tan\beta$ and small values of $\lambda$. 
Such suppression is also possible for small $\tan\beta$ and large $\lambda$ but
at the cost of additional fine-tuning of the NMSSM parameters. 

\section*{Acknowledgments}

This work is a part of the ``Implications of the Higgs boson discovery on supersymmetric extensions of the Standard Model'' project funded within the
HOMING PLUS programme of the
Foundation for Polish Science. MO and SP have been supported by National Science Centre under research grants DEC-2011/01/M/ST2/02466,
DEC-2012/04/A/ST2/00099, DEC-2012/05/B/ST2/02597.
MB has been partially supported by the MNiSW grant IP2012 030272.

\end{document}